\documentclass[a4paper,11pt]{article}
\usepackage{amsfonts,amssymb,graphicx,amsmath}
\usepackage{amscd}

\newtheorem{theorem}{Theorem} [section]

\newtheorem{lemma}[theorem]{Lemma}

\newtheorem{proposition}[theorem]{Proposition}

\newenvironment{proof}[1][Proof]{\noindent\textbf{#1.} }{\ \rule{0.5em}{0.5em}}

\def\a{\alpha}

\begin{document}

\author{M. Dror\thanks{\emph{Corresponding author}. Management Information
Systems, Eller College of Management, University of Arizona, Tucson,
Arizona, 85721. e-mail: mdror@eller.arizona.edu}, L. A. Guardiola\thanks{%
Operations Research Center, Universidad Miguel Hern\'{a}ndez, Elche, Spain.
e-mail: ana.meca@umh.es}, A. Meca$^\dag$ and J. Puerto\thanks{%
Facultad de Matem\'aticas, Universidad de Sevilla, 41012 Sevilla, Spain.
e-mail: puerto@us.es }}
\title{\textbf{Dynamic Realization Games in Newsvendor Inventory
Centralization}}
\date{\today}
\maketitle

\begin{abstract}
Consider a set $N$ of $n (>1)$ stores with single-item and single-period
nondeterministic demands like in a classic newsvendor setting with holding
and penalty costs only. Assume a risk-pooling single-warehouse centralized
inventory ordering option. Allocation of costs in the centralized inventory
ordering corresponds to modelling it as a cooperative cost game whose
players are the stores. It has been shown that when holding and penalty
costs are identical for all subsets of stores, the game based on optimal
expected costs has a non empty core (Hartman \textit{et. al.}, 2000, Muller
\textit{et. al.}, 2002). In this paper we examine a related inventory
centralization game based on demand realizations that has, in general, an
empty core even with identical penalty and holding costs (Hartman and Dror,
2005). We propose a repeated cost allocation scheme for dynamic realization
games based on allocation processes introduced by Lehrer (2002a). We prove
that the cost subsequences of the dynamic realization game process, based on
Lehrer's rules, converge almost surely to either a least square value or the
core of the expected game. We extend the above results to more general
dynamic cost games and relax the independence hypothesis of the sequence of
players' demands at different stages.

\bigskip \noindent \textbf{Key words:} Dynamic realization games, newsvedor
centralization game, cooperative game, allocation process, core, least
square value.

\end{abstract}

\section{Introduction}

Consider a situation in which a number of firms or subsidiaries of the same
firm undertake a joint project --- for instance, centralization of inventory
handling facilities. An issue of concern in joint projects is to arrive at a
cost allocation acceptable to all parties. Participation of parties in a
joint venture and the issue of allocation of its costs models such
undertaking in a cooperative game theory framework. In this paper we examine
a well known newsvendor inventory centralization setting.

\medskip

Suppose we have a set of stores that distribute a single product to their
customers. A joint project might consider filling orders from a central
facility and shipping directly to the stores' customers -- as in catalog
stores with sample merchandize. Now suppose that the demand at each of the
stores varies randomly with a given distribution function $F_i$ for store $i$
($i=1,2,\ldots, n$) and parameters specific to each store. Each store may
independently decide to participate in the centralized ordering arrangement.
When participating, the stores would share the costs of the centralized
inventory and benefit from the resulted savings.

\medskip

There have been a number of studies that examined a combined problem of
optimizing savings from a centralized inventory and allocating the savings
in a manner which maintains the cooperation of participants (Parlar, 1988,
Hartman and Dror, 1996, Anupindi and Bassok, 1999, Hartman et al. 2000,
Hartman and Dror, 2003, 2005, Slikker et al. 2005, Burer and Dror, 2006).
When logistics providers are setting up inventory and distribution
coordination, it is usually referred to as \textit{supply chain management}.
In this context Naurus and Anderson (1996), provide a number of enlightening
examples of cost cutting when inventory management is coordinated and
centralized. If the sharing of benefits is not perceived to be equitable by
the participating parties, the ``partnership" may fall apart and the overall
benefits might be lost. As Naurus and Anderson (1996) point out,
``significant hurdles stand between the idea and its implementation. To
begin with, channel members are likely to be skeptical about the rewards of
participation...".

\medskip

The focus of this study is the newsvendor problem, and we start with general
remarks regarding the demand distribution and incentives for centralization
in an infinitely repeated single period problem.

\medskip

For general demand distributions $F_i, i=1,2,\ldots,n$ and store specific
holding and penalty costs there might not be any savings from
centralization. Conditions on demand distributions are discussed in Chen and
Lin (1989) and on holding and penalty costs in Hartman and Dror (2005). In
this study we assume inventory models where the demand at each store is any
random variable having null probability of achieving negative values, and
allow for correlated stores' demands. We assume identical storage and
penalty costs for each store and in the centralized location. Eppen (1979)
was the first to show that in this case, with identical and constant holding
and penalty costs, savings always occur.

\medskip

The newsvendor inventory centralization problem examined so far in the
literature is geared to the expected value cost analysis. However,
minimizing expected centralized inventory cost might not be a very
convincing argument for centralization. A build-in cost allocation mechanism
should provide additional incentives for cooperation. That is, in each time
period the stores reflect on the actual performance of the system in
relation to the anticipated long-run expected performance. The analysis of
an on-line system cost allocation(s) performance versus the performance in
expectation is the main topic of this paper.

\medskip

The outline of the paper is as follows. Section 2 introduces two games: the
newsvendor centralization game referred to as the \textit{newsvendor
expected game} and the related \textit{newsvendor realization game}. In
addition, another related game, the \textit{dynamic newsvendor realization
game} is defined. In section 3 we discuss the repeated allocation process
introduced by Lehrer (2002a) and summarize its main findings. In section 4
we contrast the dynamic realization game with the stylized game of Lehrer
(2002a) and prove two results. That is, we prove that the diagonal
allocation sequence based on Lehrer's rule $R_1$ applied to the dynamic
realization games converges almost surely to some least square value of the
expectation game. The other main result states that any accumulation point
of the diagonal sequence of allocations based on Leherer's rule $R_2$
belongs almost surely to the core of the expected game. In section 5 we
extend the above results to allocation processes for more general dynamic
cost games. We then relax the independence hypothesis of the sequence of
players' demands at different stages, replacing it with an appropriate
strong stationarity requirement. Similar almost surely convergence results
follow.

\section{Inventory centralization in newsvendor environments}

Suppose a finite set of stores (newsvendors) that respond to a periodic
random demand (of newspapers) by ordering a certain quantity at the start of
every period. Since the demand is random, in each period a store will face
one of two cases: (1) the ordered quantity is less than the realized demand
resulting in lost profit for the store; (2) the ordered quantity exceeds the
realized demand resulting in a disposal cost for the store since the items
(the newspapers) are perishable.

\medskip

Formally, we consider a set $N=\{1,...,n\}$ of stores. The assumptions of
this model are:

\begin{enumerate}
\item Let $\left( \Omega ,\mathcal{F},P\right) $ be a given probability
space. Each store $i\in N$ faces a nonnegative random demand $x_{i},$ with
distribution function $F_{i}$ and mean $\mu _{i}$.

\item The disposal cost is $h>0$ per unit and lost profit (penalty) cost is $%
p>0$ per unit. These costs are the same for all stores and any combination
of the stores.

\item The product is ordered once at the start of each period (cannot be
reordered), and items on hand at the beginning of the period cannot be
returned. There is no order cost, and no quantity discounts or price breaks.
Both the demand distributions and the costs are common knowledge and it is a
stationary, infinitely repetitive setting.

\item The cost resulting from an initial inventory of $q$ is%
\begin{equation*}
\Psi (x,q)=\left\{
\begin{array}{cc}
h(q-x) & \text{if }q\geq x \\
p(x-q) & \text{if }q<x%
\end{array}%
\right.
\end{equation*}

\item Consider a coalition $S\subseteq N$ of stores facing a joint demand $%
x_{S}=\sum_{i\in S}x_{i}$ with distribution function $F_{S}$ and expected
value $\mu _{S}$. Assume that for all coalitions $S\subseteq N,F_{S}\in
\pounds ^{1}\left( \Omega ,\mathcal{F},P\right) $. Hence, $E\left[ \Psi
(x_{S},q)\right] <\infty $ for all $q\in \mathbb{R}$.

\medskip
\end{enumerate}

The above is a classic newsvendor setting and for all $S\subseteq N,$ we can
find a value $q_{S}$ that minimizes $E\left[ \Psi (x_{S},q)\right] ;$ $i.e.,$
$q_{S}$ is the optimal order size for $S.$

\medskip

We define the \emph{newsvendor expected game} $(N,c_{E})$ (henceforth $E$%
-game) as the TU cost game with characteristic function $c_{E}(S):=E\left[
\Psi (x_{S},q_{S})\right] $ for all $\emptyset \neq S\subseteq N.$ Notice
that for $S\subseteq N,c_{E}(S)\geq 0$ represents the optimal expected cost
of holding or shortage.

\medskip

Hartman et al. (2000) proves that the core of $E$-games is non-empty (i.e., $%
E$-games are balanced) for demands with symmetric distribution and for joint
multivariate normal demand distribution. M\"{u}ller et al. (2002)
generalizes the balancedness character of $E$-games for all possible joint
distributions of the random demands.

\medskip

Suppose now that in a given period, coalition $S\subseteq N$ decides on an
optimal order size $q_{S}$. Then, at the end of the period, each player $%
i\in S$ observes its demand realization, say $\hat{q}_{i}.$ The total demand
realization for $S$ is $\hskip1ex\hat{q}(S)=\sum_{i\in S}\hat{q}_{i}.$ Just
as for a single store there are two possibilities: (1) $\hat{q}(S)\leq q_{S}$%
, and the cost for this centralized system is $h(q_{S}-\hat{q}(S));$ (2) $%
\hat{q}(S)\geq q_{S}$ and the cost is equal to $p(\hat{q}(S)-q_{S}).$

\medskip

The \emph{newsvendor realization game} (henceforth $R$-game), $(N,c_{R}),$
is defined by $c_{R}(S):=\max \left\{ p(\hat{q}(S)-q_{S}),h(q_{S}-\hat{q}%
(S))\right\} $ for all $\emptyset \neq S\subseteq N,$ where $q_{S}$ is the
demand of $S$ in the $E$-game$.$ This non-negative game measures the actual
cost of the demand realization for every $S\subseteq N$. Hartman and Dror
(2005) shows that $R$-games are not balanced in general (i.e., the core may
be empty), by providing a realization example for joint multivariate normal
demand distribution.

\medskip

The following proposition state that $E$-games and $R$-games are related by
means of a long-term expectation property.

\begin{proposition}
Consider an $E$-game $(N,c_{E})$ and its corresponding $R$-game $(N,c_{R}).$
Then, $E\left[ c_{R}(S)\right] =c_{E}(S)$ for all $\emptyset \neq S\subseteq
N.$
\end{proposition}

\begin{proof}
Observe that $g(x)=\max \{p(x-q_{S}),h(q_{S}-x)\}$ for all $x\in \mathbb{R}$
can be rewritten as follows:

\begin{equation}  \label{eq:g_def}
g(x)=\left\{
\begin{array}{cc}
p(x-q_{S}), & x\geq q_{S} \\
h(q_{S}-x), & x<q_{S}%
\end{array}%
\right. .
\end{equation}

Thus, for all possible joint distributions of the random demands,

\begin{equation*}
E\left[ c_{R}(S)\right] =E\left[ g(\hat{q}(S))\right] =c_{E}(S).
\end{equation*}
\end{proof}

\medskip

Notice that the above property means that the long-term average cost of
coalition $S$, for repeated realizations of the actual cost game $c_{R}$, is
the same as its cost in the expected cost game $c_{E},$ provided that the
underlying individual demand distributions do not change.

\medskip

To complete this section we introduce a dynamic analysis of the newsvendor
situation; i.e., we take into account the temporal aspect of newsvendor
realization games.

\medskip

Consider the stores over any finite horizon $T$ ($T$ is a positive integer
counting the number of single periods in the finite horizon). In every
period $t$ we observe actual demands quantities $\hat{q}_{i}^{t}$, for all $%
i=1,\ldots ,n$. For a fixed $t$, those are realizations of the demand random
variables. The sequence of demand realizations faced by store $i$, $\left\{
\hat{q}_{i}^{t}\right\} _{t\geq 1}$, is ruled by the distribution function $%
F_{i},$ and we assume that $\hat{q}_{i}^{t}$ and $\hat{q}_{i}^{t^{\prime }}$
are independent for any $t\neq t^{\prime }.$ Hence $\hat{q}_{i}^{t}$ and $%
\hat{q}_{i}^{t^{\prime }}$ are independent and identically distributed
(i.i.d.) random variables for all $t\neq t^{\prime }$.

\medskip

Consider, for each store $i$, $i=1,\ldots ,n,$ the average sequence of
demand realizations $\left\{ \tilde{q}_{i}^{T}\right\} _{T\geq 1}$, where $%
\tilde{q}_{i}^{T}:=\frac{1}{T}\sum_{t=1}^{T}\hat{q}_{i}^{t}.$

\medskip

We define the \emph{dynamic newsvendor realization game} ($DR$-game) at
stage $T,(N,\tilde{c}_{R}^{T}),$ by the following characteristic function:%
\begin{equation}
\tilde{c}_{R}^{T}(S):=\max \{p(\tilde{q}^{T}(S)-q_{S}),h(q_{S}-\tilde{q}%
^{T}(S))\},
\end{equation}%
for all $\emptyset \neq S\subseteq N,$ where $\tilde{q}^{T}(S)=\sum_{i\in S}%
\tilde{q}_{i}^{T}$.

\medskip

Given a sequence of actual demand's realizations $\hat{q}=\{\hat{q}%
^T\}_{T\ge 1}$ with $\hat{q}^T=(\hat{q}_{i}^{T})_{i=1,\ldots ,n}$, $DR$%
-games ${(N,\tilde{c}_{R}^{T}(\hat{q}))}_{T\geq 1}$ are standard TU
cooperative games.

The stochastic aspect of the $DR$-games ${(N,\tilde{c}_{R}^{T}(\hat{q}))}%
_{T\geq 1}$ is ruled by the random demands from where we draw the $(\hat{q})$
realizations and it is described through the sequences of random variables $%
\{\tilde{c}_{R}^{T}(S)\}_{T\geq 1}$ for all $S\subseteq N$. Note that the
probability distribution function of the random variable $\tilde{c}_R^T(S)$
is given by:

\begin{equation*}
P[\tilde{c}_R^T(S)\leq r]=P\left[%
\begin{array}{l}
\mbox{all the outcomes of } \hat{q}_i^t, \; i=1,\ldots,n, t=1,\ldots,T %
\mbox{ such that } \\
\max \{p(\tilde{q}^{T}(S)-q_{S}),h(q_{S}-\tilde{q}^{T}(S))\}\leq r%
\end{array}%
\right],
\end{equation*}
for all $r\in \mathbb{R}$.

\medskip

Hence, the class of $DR$-games is a subclass of the class of cooperative
games with random costs. Notice that for any sequence of actual demand's
realizations $\hat{q}$, the $DR$-game at stage $T=1$ coincides with the $R$%
-game; i.e., $\tilde{c}_{R}^{1}(\hat{q})(S)=c_{R}(S) $ for all $S\subseteq
N. $

\medskip\ \

There are few models of cooperative games where the cost (or payoff) of a
coalition may be uncertain; these games are called stochastic cooperative
games. For a clear and detailed overview Suijs (2000) (see also Granot,
1977). Fern\'{a}ndez et al. (2002) introduce cooperative games with random
payoffs for the coalitions. The random payoffs are compared by means of
stochastic orders. Timmer et al. (2003, 2005) and Timmer (2006) study a
model where the stochastic value of coalitions depends on a set of actions
that every coalition can take. Several solution concepts (Core,
Shapley-like, compromise value) for all the above stochastic games have been
analyzed.

\medskip

In this paper, we focus on solutions for $DR$-games by means of dynamic
(time dependent) processes. Our analysis builds upon the work of Lehrer
(2002a,b).

\section{\label{Lehrer}Repeated allocation processes}

In real-life we may expect players to monitor their costs of inventory
centralization one period at a time and, accordingly, form an
\textquotedblleft opinion/response" regarding the fairness of their
respective cost allocations. From now on, we focus on a repeated allocation
process.

\medskip

Lehrer (2002a) describes four allocation rules in a stylized cooperative
game repeated an infinite number of times. At each time period the same game
is played with a fixed budget of size $B$ that has to be distributed among
the players in a finite set $N$. The game has its characteristic function $v$
with the interpretation that $v(S)$ represents the needs of coalition $%
S\subseteq N$. The allocation at time period $t,t=0,1,2,\ldots $ is a vector
$a_{t}=(a_{t}^{i})$, where $a_{t}^{i}$ is the portion of the budget $B$
allocated to player $i\in N$ at time $t$. An allocation rule determines the
allocation at time $T,a_{T},$\ as a function of $v$ and of all $a_{t},t<T$.
The sequence of allocations $(a_{t})_{t\geq 1}$ is the allocation process
induced by the rule.

\medskip

The empirical distribution of the budget among the players is, at any stage,
an allocation of the budget. We focus on the first two types of processes
introduced by Lehrer, to be implemented later on (see next section) the $DR$%
-games at each stage $T$. The first type of process is based on the idea
that giving a budget to a player increases the total well-being of the
entire group. The player whose marginal contribution to this well-being is
maximal will receive the budget. It is shown that any process of this type
generates allocations that converge to some least square value (introduced
by Ruiz et al., 1998). The second type of allocation process is defined
inductively and a player whose weighted actual surplus is nonnegative is
chosen and given the entire budget. This process converges either to the
core of the game, when the game is balanced, or to the least core.
Convergence in this context means that the distance between the core (or the
least core) and the empirical sequence of allocations shrinks to zero. The
proofs related to the above two types of allocation processes rely on a
geometric principle that lies behind Blackwell's approachability result
(Blackwell, 1956, and Lehrer, 2002b).

\medskip

Formally, let $N$\ be a finite set of players, where the number of players, $%
|N|,$\ is $n >1.$\ Consider a normalized cooperative game $v$. Let $%
A=\{(a_{1},a_{2},\ldots ,a_{n})\in \mathbb{R}^{n}: \sum_{i=1}^{n}a_{i}=v(N)=1%
\text{ and }a_{i}\geq v(\{i\})\text{ for all }i\in N \}$\ be the set of
allocations. An allocation rule $R$\ is a function $R:\cup _{t=0}^{\infty
}A^{t}\rightarrow A,$\ where $A^{t}$\ is the Cartesian product of $A$\ with
itself $t$\ times and $A^{0}$\ is a singleton that represents the empty
history of allocations. As mentioned above, an allocation rule at time $T$\
is a function that determines the allocation $a_{T}$\ as a function of past
allocations $a_{t},t<T.$\ In fact, it induces a sequence $a_{1},a_{2},...$
of allocations in $A$ as follows: $a_{1}$\ is the first allocation $R$\
prescribes, $a_{2}=R(a_{1}),a_{3}=R(a_{1},a_{2}),$\ etc. This sequence is
called the allocation sequence induced by $R.$

\medskip

Below we summarize the first two Lehrer's allocation rules by means of the
corresponding allocation process.

\begin{enumerate}
\item \emph{Processes that converge to the least square value ($R_1$
allocation rule)}

For any time $t$, denote by $\overline{a}_{t}$\ the historical distribution
of the budget up to time $t.$\ That is, $\overline{a}_{t}^{i}$\ is the
frequency of the stages up to time $t,$\ where player $i\in N$\ received the
budget. For any $S\subseteq N,$\ let $\overline{a}_{t}(S)$\ be $\sum_{i\in S}%
\overline{a}_{t}^{i}.$\ A coalition is chosen randomly according to the
probability distribution $(\alpha _{S})_{S\subseteq N}.$\ At time $t+1$\ the
coalition $S$\ is assigned a weight proportional to the excess corresponding
to the allocation $\overline{a}_{t},\overline{a}_{t}(S)-v(S).$\ At any time
a player whose contribution to the expected weighted welfare of society $%
\sum_{S\subseteq N}\alpha _{S}\left( \overline{a}_{t}(S)-v(S)\right) \left(
\mathbb{I}_{i\in S}-v(S)\right) $, is maximal, is chosen (where $\mathbb{I}%
_{i\in S}-v(S)$\ is $1-v(S)$\ if $i\in S$\ and $-v(S)$\ otherwise). This
player receives the entire budget and is denoted player $i_{t+1}.$\ Denote
by $R_{1}$\ the allocation rule induced by the above process (see Lehrer
(2002a) for further details).

The following theorem shows that this $R_{1}$\ allocation rule generates an
allocation sequence that converges to some least square value. Then, one can
select for instance the Shapley value as a special case.

\begin{theorem}
\label{lehert1}\emph{(Lehrer, 2002a)}. Let $a_{1},a_{2},...$\ be the
allocation sequence induced by $R_{1}.$\ Then, $\overline{a}_{t}$\ converges
to the least square value of the game that corresponds to the weights $%
\alpha _{S},S\subseteq N.$
\end{theorem}

\item \emph{Processes that converge to the core and to the least core ($R_2$
allocation rule)}

Let $v$ be a balanced game. Let $S_{1},...,S_{k},$ $k=2^{n}-1$, be the list
of all non-empty coalitions of $N.$ Denote by $y_{i}$ the vector in $\mathbb{%
R}^{k}$ whose $l^{th}$ coordinate is $\mathbb{I}_{i\in S_{l}}-v(S_{l}).$ Two
sequences are defined: the allocation process $a_{1},a_{2},...$ of vectors
in $\mathbb{R}^{n},$ by means of vectors of the standard basis of $\mathbb{R}%
^{n},$ and an auxiliary sequence $z_{1},z_{2},...$ of vectors in $\mathbb{R}%
^{k},$ by means of vectors $y_{i}$ in $\mathbb{R}^{k}$ (see Lehrer (2002a)
for further details). The figure $\overline{z}_{t}^{l}$ measures the
historical average surplus of the coalition $S_{l}$ up to stage $t.$ At this
stage the coalitions are weighted with respect to these surpluses: those
coalitions with a positive surplus are neglected while the other coalitions
are assigned a weight proportional to their (negative) surplus (i.e., for
such a coalition, say, $S_{l},$ the weight is $-\min (\overline{z}_{t}^{l},0)
$). Then, a player $i$ whose weighted actual surplus (i.e., $%
\sum_{l=1}^{k}[-\min (\overline{z}_{t}^{l},0)][\mathbb{I}_{i\in
S_{l}}-v(S_{l}])$ is non-negative is chosen and is given the entire budget $%
v(N).$ The vector $\overline{a}_{t}$ is the historical distribution of the
budget up to time $t.$ Let $R_{2}$\ be the allocation rule induced by the
above process.

The following theorem shows that any limit point of the corresponding
allocation sequence $\overline{a}_{t}$ is in the core.

\begin{theorem}
\label{lehert2}\emph{(Lehrer, 2002a)}. Let $a_{1},a_{2},...$\ be the
allocation sequence induced by $R_{2}.$\ Then, $\overline{a}_{t}$\ converges
to the core of the game. That is, any accumulation point of the sequence $%
\overline{a}_{t}$ is in the core.
\end{theorem}

Notice that in constructing the allocation process that converges to the
core we assumed that the game is balanced. As noted in Lehrer (2002a), rule $%
R_{2}$ can be modified in the case of an empty core to obtain an allocation
sequence which converges to a point in the least core (the intersection of
all non-empty $\epsilon $-cores).
\end{enumerate}

\section[The dynamic newsvendor realization game]{Allocation processes for
the dynamic newsvendor realization game}

First, we examine the differences between the stylized \textquotedblleft
budget" game presented in Lehrer (2002a) and the dynamic newsvendor
realization game introduced in this paper. From now on we refer to the game
described in Lehrer (2002a) as the $Le$-game.

\begin{enumerate}
\item The $Le$-game has a fixed budget $B$ in each time period. The $DR$%
-game has different values obtained by randomly drown demand values for the
different players in the expected value newsvendor centralization problem.
Thus the $DR$-game is not normalized, or alternatively has to be normalized
at each time period.

\item The $DR$-game is defined in reference to the expected value newsvendor
centralization game known to posses a nonempty core where as the $Le$-game
is a stand alone game with or without an empty core. The core of a $DR$%
-game, regarded as a stochastic cooperative game, is a random set and needs
to be explicitly defined.
\end{enumerate}

In this section, we prove that there exist Lehrer's allocation processes
(rules $R_1$ or $R_2$) applied to the $DR$-games at each stage $T$, that
converge almost surely (that is, the set of outcomes where it does not
converges has null probability) either to some least square value or to the
core of $E$-games.

\medskip

We start proving a technical lemma that will be useful in our analysis.
Previously, given any TU game $(N,c)$, we recall that the least square value
(see Ruiz et al., 1998) for a weight function $\alpha =(\alpha
_{S})_{S\subseteq N}$ is:
\begin{equation*}
LS^{\alpha }(c)=(LS_{1}^{\alpha }(c),\ldots ,LS_{n}^{\alpha }(c)),
\end{equation*}%
where
\begin{equation*}
LS_{i}^{\alpha }(c)=\displaystyle\frac{c(N)}{n}+\frac{1}{n\beta }\Big[%
\sum_{S:i\in S}(n-s)\alpha (S)c(S)-\sum_{S:i\not\in S}s\alpha (S)c(S)\Big]
\end{equation*}%
and $\beta =\displaystyle\sum_{s=1}^{n-1}\alpha (S){\binom{{n-2}}{{s-1}}}.$

\medskip

Notice that the least square value for a weight function $\alpha$ can easily
be extended to stochastic cooperative games; in particular to DR-games.
Indeed, take the continuous functions $f_{i}: \mathbb{R}^{2^{n}-1}
\longrightarrow \mathbb{R}$ defined by
\begin{equation*}
\begin{array}{rccl}
f_{i}(x)=\frac{g(x_{N})}{n}+\frac{1}{n\beta }\Big[\sum_{S:i\in S}(n-s)\alpha
(S)g(x_{S})-\sum_{S:i\not\in S}s\alpha (S)g(x_{S})\Big], &  &  &
\end{array}%
\end{equation*}%
for all $x=(x_{S})_{S\subseteq N} \in \mathbb{R}^{2^{n}-1}$. (Recall the $g$
was introduced in \eqref{eq:g_def}.) Then, for all $i=1,\ldots ,n$ and all $%
T\geq 1$,

\begin{equation}  \label{fi}
LS_{i}^{\alpha }(\tilde{c}_{R}^{T})=f_{i}((\tilde{c}_{R}^{T}(S))_{S\subseteq
N})
\end{equation}

\noindent is a random variable with values $LS_{i}^{\alpha }(\tilde{c}%
_{R}^{T}(\hat{q}))=f_{i}((\tilde{c}_{R}^{T}(\hat{q})(S))_{S\subseteq N})$,
for any sequence of actual demand's realizations $\hat{q}$.

\begin{lemma}
\label{le1} For any $S\subseteq N$ and any weight function $\alpha $ (that
does not depend on $T$),
\begin{eqnarray}
\tilde{c}_{R}^{T}(S) &\overset{a.s.}{\underset{T\rightarrow \infty }{%
\longrightarrow }}&c_{E}(S).  \label{le1:ass1} \\
LS_{i}^{\alpha }(\tilde{c}_{R}^{T}) &\overset{a.s.}{\underset{T\rightarrow
\infty }{\longrightarrow }}&LS_{i}^{\alpha }(c_{E})\quad \forall i\in N.
\label{le1:ass2}
\end{eqnarray}
\end{lemma}

\begin{proof}
Consider the continuous function
\begin{equation*}
\begin{array}{rlcl}
g: & \mathbb{R} & \rightarrow  & \mathbb{R} \\
& x & \rightarrow  & g(x)=\max \{p(x-q_{S}),h(q_{S}-x)\}.%
\end{array}%
\end{equation*}%
Recall that $\hat{q}^{t}(S)=\sum_{i\in S}\hat{q}_{i}^{t}$ and consider the
sequence $\{\hat{q}^{t}(S)\}_{t\geq 1}.$ It is clear that the random
variables in the sequence are i.i.d. with mean value $\mu _{S}$. Therefore,
by the Strong Law of Large Numbers,%
\begin{equation*}
\tilde{q}^{T}(S)=\displaystyle\frac{1}{T}\sum_{t=1}^{T}\hat{q}^{t}(S)\overset%
{a.s.}{\underset{T\rightarrow \infty }{\longrightarrow }}\mu _{S}.
\end{equation*}%
Hence, by the continuity of $g$,
\begin{equation*}
g(\tilde{q}^{T}(S))=\tilde{c}_{R}^{T}(S)\overset{a.s.}{\underset{%
T\rightarrow \infty }{\longrightarrow }}g(\mu _{S})=c_{E}(S)
\end{equation*}%
which proves \eqref{le1:ass1}.

Applying \eqref{le1:ass1} we have
\begin{equation*}
LS_{i}^{\alpha }(\tilde{c}_{R}^{T})=f_{i}((\tilde{c}_{R}^{T}(S))_{S\subseteq
N})\overset{a.s.}{\underset{T\rightarrow \infty }{\longrightarrow }}%
f_{i}((c_{E}(S))_{S\subseteq N})=LS_{i}^{\alpha }(c_{E}).
\end{equation*}%
This proves \eqref{le1:ass2}.
\end{proof}

\medskip

Our approximation process consists of applying Lehrer's allocation rules $%
R_{1}$ and $R_{2}$ (described in section \ref{Lehrer}) to each one of the
our $DR-$games, at each stage $T$, and then to derive allocation rule
properties of the repeated realization process.

\medskip

Assume the we are given a generic game $(N,c)$. Let $\{{a}%
^{R_i}(c)(l)\}_{l\ge 1}$ denote Lehrer's allocation sequence induced by $R_i$
$i=1,2$. (By ${a}^{R_i}(c)(l)$ we refer to the $l$-th element in the
corresponding sequence.) Let $\{\bar{a}^{R_1}(c)(l)\}_{l\ge 1}$ and $\{\bar{a%
}^{R_2}(c)(l)\}_{l\ge 1}$ be the allocation schemes converging to some least
square value with weight function $\alpha $ (the former) and converging to
an element of the least core (the latter). \medskip

Notice that Lehrer's allocation sequence induced by $R_i$ $i=1,2$, can also
be extended to stochastic cooperative games, in particular to $DR-$games.
Indeed, Lehrer's allocation scheme applied over a $DR-$game $(N,\tilde{c}%
_{R}^{T})$ is a random variable $\bar{a}^{R_{i}}(\tilde{c}_{R}^{T})$, with
values $\bar{a}^{R_{i}}(\tilde{c}_{R}^{T}(\hat{q}))$ for any sequence of
actual demand's realizations $\hat{q}$.

\medskip

Our first theorem ensures the convergence of Lehrer's $\{\bar{a}%
^{R_1}(c)(l)\}_{l\ge 1}$ allocation scheme, applied over the $DR-$games, to
some least square value of $E- $games. In this case, we obtain that for any
sequence of actual demand realizations the sequence of diagonal steps $\{%
\bar{a}^{R_{1}}(\tilde{c}_{R}^{T})(T)\}$ (i.e. the one that chooses the $T$%
-th replication of Lehrer's $R_1$-approach at stage $T$ of the realization
game) converges almost surely to some least square value of the $E-$games.

\begin{theorem}
\label{t:LS} Let $(N,c_{E})$ be a newsvendor expected game. The diagonal
sequence $\{\bar{a}^{R_{1}}(\tilde{c}_{R}^{T})(T)\}_{T\geq 1}$ converges
almost surely to $LS^{\alpha }(c_{E})$.
\end{theorem}

\begin{proof}
First of all, we note that according to Theorem \ref{lehert1}, for any
sequence of actual demand's realization $\hat{q}$,
\begin{equation*}
\bar{a}^{R_{1}}(\tilde{c}_{R}^{T}(\hat{q}))(l)\overset{pointwise}{\underset{%
l\rightarrow \infty }{\longrightarrow }}LS^{\alpha }(\tilde{c}_{R}^{T}(\hat{q%
})),\forall T\geq 1.
\end{equation*}

Therefore, applying the pointwise convergence of the above process together
with \eqref{le1:ass2} we have the diagram:

\begin{eqnarray*}
\begin{CD} LS^{\a}(\tilde{c}_R^T) @<{pointwise}<{l\rightarrow \infty}<
\bar{a}^{R_{1}}(\tilde{c}_R^T)(l)\\ @V{T\rightarrow \infty}V{a.s.}V \\
LS^{\a}(c_E). \end{CD}
\end{eqnarray*}

Hence, taking any infinite subsequence with increasing indexes in $(l,T)$ we
obtain almost sure convergence to $LS^{\alpha }(c_{E})$. In particular,
following the diagonal sequence, namely taking indexes $(T,T)$, $T\geq 1$,
we get the result in the theorem.
\end{proof}

\medskip

Our next result explains the approachability of $\{\bar{a}%
^{R_{2}}(c)(l)\}_{l\geq 1}$ allocation scheme, applied to $DR-$games, to the
core of $E$-games. In this case, we prove that any accumulation point of the
sequence of diagonal steps $\{\bar{a}^{R_{2}}(\tilde{c}_{R}^{T})(T)\}$
converges almost surely to a point in the core of $E$-games.

\medskip

The meaning of $core(c_E)$ is clear:
\begin{equation*}
core(c_E)=\{x\in \mathbb{R}^n: \sum_{i\in S} x_i\le c_E(S),\; \forall
S\subset N,\; \sum_{i\in N} x_i=c_E(N)\}.
\end{equation*}

\medskip

The same meaning is applicable to the core of the realization game defined
on a sequence of actual demand's realizations $\hat{q}=\{\hat{q}^t\}_{t\le
T} $ at any stage $T$; i.e. $core(\tilde{c}_R^T(\hat{q}))$. Analogously, the
$\varepsilon$-core of that game for any $\varepsilon>0$ is:
\begin{equation*}
core(\tilde{c}_R^T(\hat{q}),\varepsilon)=\{x\in \mathbb{R}^n: x(S)\le \tilde{%
c}_R^T(\hat{q})(S)+\varepsilon,\; S\subset N;\;x(N)= \tilde{c}_R^T(\hat{q}%
)(N)\},
\end{equation*}
and the least core is:
\begin{equation*}
Lcore(\tilde{c}_R^T(\hat{q}))=\bigcap_{core(\tilde{c}_R^T(\hat{q}%
),\varepsilon)\neq \emptyset} core(\tilde{c}_R^T(\hat{q}),\varepsilon).
\end{equation*}

\medskip

However, the meaning of $core(\tilde{c}_{R}^{T})$ and $core(\tilde{c}%
_{R}^{T},\varepsilon )$ is not clear since $\tilde{c}_{R}^{T}$ is a
stochastic cooperative game at each stage T. Therefore, we must first define
these sets.

\medskip

First, we have to extend the concept of efficiency. Note that when $\tilde{c}%
_{R}^{T}(N)$ is an absolutely continuous random variable, requiring $\tilde{c%
}_{R}^{T}(N)=x(N)$ would imply a zero probability event. Thus, inducing the
concept of core to be a set with null probability. To overcome this
difficulty, we define efficiency through a significance level around the
average value of the random variable.

\medskip

Given a random variable $Y$ with $E(Y)=\bar{y}$ and a significance level $%
\beta$, $0\leq \beta \leq 1$, let $\phi (\beta )=\inf \{\phi ^{\prime }:P[|Y-%
\bar{y}|\leq \phi ^{\prime }]\geq \beta \}$. We say that a vector $x$ is $%
\phi (\beta )-$efficient if $|x(N)-\bar{y}|\leq \phi (\beta )$.

\medskip

In our setting $core(\tilde{c}_{R}^{T})$ and $core(\tilde{c}%
_{R}^{T},\varepsilon )$ are random sets. Therefore, to define their meaning
we have to state the significance level $\beta _{T}$ of that efficiency,
which in turns induces the values $\phi _{T}(\beta _{T})$. For simplicity,
we denote those values $\phi _{T}(\beta _{T})$ as $\phi _{T}$, when no
confusion is possible. Then, the probability of an allocation $x$ to be in
those cores is given as:
\begin{equation}
P[x\in core(\tilde{c}_{R}^{T})]=P\left[ x(S)\leq \tilde{c}%
_{R}^{T}(S),\forall S\subset N;|x(N)-E(\tilde{c}_{R}^{T}(N))|\leq \phi _{T}%
\right] ,
\end{equation}%
\begin{equation}
P[x\in core(\tilde{c}_{R}^{T},\varepsilon )]=P\left[ x(S)\leq \tilde{c}%
_{R}^{T}(S)+\varepsilon ,\;\forall S\subset N;|x(N)-E(\tilde{c}%
_{R}^{T}(N))|\leq \phi _{T}\right] .  \label{def:coreR}
\end{equation}

\medskip

Note that setting $\beta_T=1$ for all $T$, provided that $\phi_T$ is not
identically equal to $+\infty$, we have by \eqref{le1:ass1} that $\phi _{T}%
\underset{T\rightarrow \infty }{\rightarrow }0 $.

\medskip

Then, $x\in core(\tilde{c}_{R}^{T})$ or $core(\tilde{c}_{R}^{T},\varepsilon
) $ almost surely if and only if $P[x\in core(\tilde{c}_{R}^{T})]=1$ or $%
P[x\in core(\tilde{c}_{R}^{T},\varepsilon )]=1$, respectively$.$

\begin{theorem}
\label{t:core} Let $(N,c_{E})$ be a newsvendor expected game. Then any
accumulation point of the sequence $\{\bar{a}^{R_{2}}(\tilde{c}%
_{R}^{T})(T)\}_{T\geq 1}$ belongs to $core(c_{E})$ almost surely.
\end{theorem}

\begin{proof}
First, we prove that there exist $\widehat{x}\in \mathbb{R}^{n}$ such that $|%
\widehat{x}(N)-E(\tilde{c}_{R}^{T}(N))|\leq \phi _{T}$ and a sequence $%
\{\varepsilon _{T}\}_{T\geq 1}$, which converges almost surely to zero, such
that $\widehat{x}\in core(\tilde{c}_{R}^{T},\varepsilon _{T})$ a.s.

\medskip

Define $\varepsilon _{T}(S):=c_{E}(S)-\tilde{c}_{R}^{T}(S),$ for all $%
S\subset N$ and let $\{\phi_T\}_T$ any sequence converging to 0. By Lemma %
\ref{le1}, $\varepsilon _{T}(S)\overset{a.s.}{\underset{T\rightarrow \infty }%
{\longrightarrow }}0$ for all $S\subset N.$ Let $\varepsilon
_{T}:=\max_{S\subset N}\left\{ \left\vert \varepsilon _{T}(S)\right\vert
\right\}$ for all $T\geq 1.$ Then, for all $\delta >0$, there exists $%
T(\delta )$ such that for all$\ T>T(\delta ),\; \varepsilon _{T}<\delta $
almost surely.

\medskip

Suppose that for all $x\in \mathbb{R}^{n}$, such that $|x(N)-E(\tilde{c}%
_{R}^{T}(N))|\leq \phi _{T}$ ($\phi _{T}$ induced by the significance level $%
\beta _{T}$), $P[x\notin core(\tilde{c}_{R}^{T},\varepsilon _{T})]>0.$ Then,

\begin{equation}
P\left[ x(S)>\tilde{c}_{R}^{T}(S)+\varepsilon _{T},\text{ for some }S\subset
N\right]>0.  \label{primera}
\end{equation}

\medskip

Now, since for any $S\subset N$, the condition $x(S)>\tilde{c}%
_{R}^{T}(S)+\varepsilon _{T}$ implies $x(S)>c_{E}(S)$, we obtain by (\ref%
{primera})\ that

\begin{equation*}
P\left[ x(S)>c_{E}(S),\text{ for some }S\subset N\right] >0,
\end{equation*}%
which is a contradiction, since taking $x^{\ast }\in core(c_{E}),$%
\begin{equation*}
P[x^{\ast }(S)>c_{E}(S),\text{ for some }S\subset N]=0.
\end{equation*}

\medskip

Hence, we conclude that there exists $\widehat{x}\in \mathbb{R}^{n}$ such
that $|\widehat{x}(N)-E(\tilde{c}_{R}^{T}(N))|\leq \phi _{T}$ satisfying $P[%
\widehat{x}\in core(\tilde{c}_{R}^{T},\varepsilon _{T})]=1.$ \bigskip

Next, we prove that for any realization of actual demands at stage $T $, $%
\hat{q}^{1},\ldots ,\hat{q}^{T},$ any accumulation point $x^{T}$ of Lehrer's
$R_{2}$ repeated allocation process, constructed on the game $(N,\tilde{c}%
_{R}^{T}(\hat{q}))$, satisfies $P[x^{T}\in core(\tilde{c}_{R}^{T},%
\varepsilon _{T})]=1$. \newline

Indeed,
\begin{equation*}
P[x^{T}\in core(\tilde{c}_{R}^{T},\varepsilon _{T})]=P\left[ \left\{ \hat{q}%
\left/
\begin{array}{l}
x^{T}(\hat{q})(S)\leq \tilde{c}_{R}^{T}(\hat{q})(S)+\varepsilon _{T}(\hat{q}%
),\forall S\subset N; \\
|x^{T}(\hat{q})(N)-E(\tilde{c}_{R}^{T}(\hat{q})(N))|\leq \phi _{T}%
\end{array}%
\right. \right\} \right]
\end{equation*}

that in turns equals 1, since Theorem 3.2 ensures that for any realization $%
\hat{q},$ $x^{T}(\hat{q})\in core(\tilde{c}_{R}^{T}(\hat{q}),\varepsilon
_{T}(\hat{q}))$ provided that this set is not empty.

\medskip

Then, we prove that for any accumulation point $\mathrm{x}$ of a sequence $%
\{x^{T}\}_{T\geq 1}$ such that $x^{T}\in core(\tilde{c}_{R}^{T},\varepsilon
_{T})$ a.s. for all $T>1,$ one has that $\mathrm{x}\in core(c_{E})$ almost
surely.

\medskip

By Lemma \ref{le1}, $\tilde{c}_{R}^{T}(S)\overset{a.s.}{\underset{%
T\rightarrow \infty }{\longrightarrow }}c_{E}(S)$ for all $S$. Thus, for any
$\delta >0$ small enough, there exists $T(\delta )$ such that for all $%
T>T(\delta )$: $\tilde{c}_{R}^{T}(S)<c_{E}(S)+\delta $ almost surely. Since $%
x^{T}\in core(\tilde{c}_{R}^{T},\varepsilon _{T})$ almost surely, it follows
that:

\medskip

\begin{equation*}
\begin{array}{l}
x^{T}(S)\leq \tilde{c}_{R}^{T}(S)+\varepsilon _{T}<c_{E}(S)+\delta
+\varepsilon _{T}, \mbox{ a.s. and for all } S\subset N, \\
\phi _{T}\geq |x^{T}(N)-E(\tilde{c}_{R}^{T}(N))|,%
\end{array}%
\end{equation*}
for all $T>T(\delta )$.

\medskip

Hence, as $\delta \rightarrow 0$ we have that $T(\delta )\rightarrow \infty ,
$ and the accumulation point $\mathrm{x}$ satisfies:
\begin{equation}
\mathrm{x}(S)\leq c_{E}(S)\mbox{ for all }S\subset N,\mbox{ a.s.}\text{ and }%
|\mathrm{x}(N)-c_{E}(N)|\leq 0,  \label{eq:5}
\end{equation}%
since $\varepsilon _{T}(S)\overset{a.s.}{\underset{T\rightarrow \infty }{%
\longrightarrow }}0$ for all $S\subseteq N$ and $\phi _{T}\underset{%
T\rightarrow \infty }{\rightarrow }0.$

\medskip

Thus, the sequence $\{\bar{a}^{R_{2}}(\tilde{c}_{R}^{T})(T)\}_{T\geq 1}$
satisfies $\bar{a}^{R_{2}}(\tilde{c}_{R}^{T})(T)\in core(\tilde{c}%
_{R}^{T},\varepsilon _{T})$ almost surely for all $T>1$. Hence, applying %
\eqref{eq:5}, any accumulation point of $\{\bar{a}^{R_{2}}(\tilde{c}%
_{R}^{T})(T)\}$ belongs to $core(c_{E})$ almost surely.
\end{proof}

\section{Final comments and remarks}

\subsection{Allocation processes for dynamic cost games}

In the previous section we present two allocation processes for dynamic
newsvendor realization games. Each of them was based on applying Lehrer's
allocation processes induced by rules $R_1$ and $R_2$, respectively, to
DR-games at each stage $T\geq 1$.

\medskip

In this subsection we extend all the above results to a more general
framework. We state two allocation processes for general dynamic cost games,
which are also based on applying Lehrer's allocation processes (rules $R_1$
or $R_2$).

\medskip

Let $(N,c)$ be a non-negative balanced TU cost game. Let $%
\{(N,c^{t})\}_{t\geq 1}$ ($t$ is no longer an index for a time period like
in a newsvendor game but simply an index of a game in a sequence) be a
sequence of non-negative stochastic cooperative games such that
\begin{equation*}
c^{t}(S)\overset{a.s.}{\underset{t\rightarrow \infty }{\longrightarrow }}%
c(S), \text{ for any } S\subseteq N.
\end{equation*}

It is clear that for a given scenario $\hat{q}$ of the sequence of
stochastic cooperative games (that is, a realization of the stochastic
behavior), $(N,c^{t}(\hat{q}))_{t\geq 1}$ are standard TU games. Therefore,
we can apply Lehrer's $R_{i}$-allocation schemes, $i=1,2,$ to each of them.

\medskip

Then we can obtain similar results to those in Theorem \ref{t:LS} and
Theorem \ref{t:core}.

\begin{theorem}
Let $(N,c)$ be a non-negative balanced TU cost game and $\{(N,c^{t})\}_{t%
\geq 1}$ a sequence of non-negative stochastic cooperative games satisfying
that $\displaystyle c^t(S)\overset{a.s.}{\longrightarrow} c(S)$, for all $%
S\subset N$. Then, the diagonal sequence $\{\bar{a}^{R_{1}}(c^{t})(t)\}_{t%
\geq 1}$ satisfies:
\begin{equation*}
\displaystyle\bar{a}^{R_{1}}(c^{t})(t)\overset{a.s.}{\underset{t\rightarrow
\infty }{\longrightarrow }}LS^{\alpha }(c).
\end{equation*}
\end{theorem}

\begin{theorem}
Let $(N,c)$ be a non-negative balanced TU cost game and $\{(N,c^{t})\}_{t%
\geq 1}$ a sequence of non-negative stochastic cooperative games satisfying
that $\displaystyle c^t(S)\overset{a.s.}{\longrightarrow} c(S)$, for all $%
S\subset N$. Then, any accumulation point of the sequence $\{\bar{a}%
^{R_{2}}(c^{t})(t)\}_{t\geq 1}$ belongs to $core(c)$ almost surely.
\end{theorem}

\subsection{Removing the independence hypothesis}

The approachability results for the $DR$-games can be further extended
removing the independence hypothesis of the sequence of players' demand at
different stages. Instead, we will require the stochastic processes $\{\hat{q%
}_i^t\}_{t\ge 1}$ to be strongly stationary for any $i\in N$. Under this
hypothesis the sequences $\{\hat{q}^t(S)\}_{t\ge 1}$ inherit the same
character (strongly stationary) and by the ergodic theorem (see Feller,
1966) the sequences $\{\tilde{q}^T(S)\}_{T\ge 1}$, where $\tilde{q}^T(S)=%
\frac{1}{T}\sum_{t=1}^T \hat{q}^t(S)$, converge almost surely to a random
variable, $Y_S$, satisfying $E[Y_S]=\mu_S$ for any $S\subseteq N$.

\medskip

Under this hypothesis one can extend, \textit{mutatis mutandis}, Lemma \ref%
{le1} which turns out to be:

\begin{lemma}
\label{le2} For any $S\subseteq N$ and any weight function $\alpha $ (which
does not depend on $T$),
\begin{eqnarray}
\tilde{c}_{R}^{T}(S) &\overset{a.s.}{\underset{T\rightarrow \infty }{%
\longrightarrow }}&Y_{S}\hspace*{3mm}\mbox{
where }E[Y_{S}]=c_{E}(S).  \label{le2:ass1} \\
LS_{i}^{\alpha }(\tilde{c}_{R}^{T}) &\overset{a.s.}{\underset{T\rightarrow
\infty }{\longrightarrow }}&Y_{LS_{i}}\mbox{ where }E[Y_{LS_{i}}]=LS_{i}^{%
\alpha }(c_{E})\quad \forall i\in N.  \label{le2:ass2}
\end{eqnarray}
\end{lemma}

Using this lemma we get similar results to Theorem \ref{t:LS}. \bigskip

\begin{theorem}
\label{t:ss_LS} Let $(N,c_{E})$ be a newsvendor expected game. The diagonal
sequence $\{\bar{a}^{R_{1}}(\tilde{c}_{R}^{T})(T)\}_{T\geq 1}$ satisfies:
\begin{equation*}
\displaystyle\bar{a}^{R_{1}}(\tilde{c}_{R}^{T})(T)\overset{a.s.}{\underset{%
T\rightarrow \infty }{\longrightarrow }}Y_{LS_{i}},\mbox{ where }%
E[Y_{LS_{i}}]=LS_{i}^{\alpha }(c_{E}),\quad \forall i=1,...,n.
\end{equation*}
\end{theorem}

The extension of Theorem \ref{t:core} seems to be more involved and needs
further investigation. The main difference is that the sequence of
characteristic functions converges now to a random vector $%
Y=\{Y_{S}\}_{S\subseteq N}$ and therefore the limit defines a random set $%
core(Y)$. The probability of an allocation to belong to $core(Y)$ is given
as:
\begin{equation}
P[x\in core(Y)]=P\left[
\begin{array}{l}
x(S)\leq Y_{S},\forall S\subset N;|x(N)-E(Y_{N})|\leq \phi _{Y}(\beta )%
\end{array}%
\right] ,
\end{equation}%
for a significance level of efficiency $\beta$. Analogously, we introduce $%
core(Y,\varepsilon )$ as the random set defined by
\begin{equation}
P[x\in core(Y,\varepsilon )]=P\left[ x(S)\leq Y_{S}+\varepsilon ,\;\forall
S\subset N;|x(N)-E(Y_{N})|\leq \phi _{Y}(\beta )\right] ,  \label{def:LcoreY}
\end{equation}

where
\begin{equation*}
\phi _{Y}(\beta ):=\inf \left\{ \delta \left/ P[|E(Y_{N})-Y_{N}|\leq \delta
]\geq \beta \right. \right\} .
\end{equation*}

\begin{theorem}
\label{t:coreY} Suppose that $Y_{S}\geq 0$ for all $S\subseteq N$, $\phi
_{Y}(1)<+\infty $ and for some $T^{\prime }>1$ there exists $x\in core(Y)$
almost surely satisfying $|x(N)-E(\tilde{c}_{R}^{T}(N))|\leq \phi _{T}$ for
all $T\geq T^{\prime }$. Then, any accumulation point $\bar{x}$ of the
sequence $\{\bar{a}^{R_{2}}(\tilde{c}_{R}^{T})(T)\}_{T\geq 1}$ belongs to $%
core(Y)$ almost surely.
\end{theorem}

\begin{proof}
First, we prove that there exist $\widehat{x}\in \mathbb{R}^{n}$ such that $|%
\widehat{x}(N)-E(\tilde{c}_{R}^{T}(N))|\leq \phi _{T}$ and a sequence $%
\{\varepsilon _{T}\}_{T\geq 1}$ which converges almost surely to zero such
that $P[\widehat{x}\in core(\tilde{c}_{R}^{T},\varepsilon _{T})]=1.$

\medskip

Define $\varepsilon _{T}(S):=Y_{S}-\tilde{c}_{R}^{T}(S),$ for all $S\subset N
$ and let $\{\phi _{T}\}_{T}$ be a sequence converging to 0. By Lemma \ref%
{le1}, $\varepsilon _{T}(S)\overset{a.s.}{\underset{T\rightarrow \infty }{%
\longrightarrow }}0$ for all $S\subseteq N.$ Set $\varepsilon
_{T}:=\max_{S\subseteq N}\left\{ \left\vert \varepsilon _{T}(S)\right\vert
\right\} ,$ for all $T\geq 1.$ Then, for all $\delta >0,\exists T$ $(\delta )
$ such that for all $\ T>T$ $(\delta ),$ $\varepsilon _{T}<\delta $ almost
surely.

\medskip

Suppose that for all $x\in \mathbb{R}^{n}$ such that $|x(N)-E(\tilde{c}%
_{R}^{T}(N))|\leq \phi _{T}$ (being $\phi _{T}$ the threshold induced by the
efficiency level $\beta _{T}$), $P[x\notin core(\tilde{c}_{R}^{T},%
\varepsilon _{T})]>0.$

\medskip

Then
\begin{equation}
P\left[ x(S)>\tilde{c}_{R}^{T}(S)+\varepsilon _{T},\text{ for some }S\subset
N\right]>0.  \label{segunda}
\end{equation}

\medskip

Now taking into account that $x(S)>\tilde{c}_{R}^{T}(S)+\varepsilon _{T}$
a.s. implies $x(S)>Y_{S}$ a.s., we obtain by (\ref{segunda}) that

\begin{equation*}
P\left[ x(S)>Y_{S},\text{ for some }S\subset N\right] >0,
\end{equation*}%
which is a contradiction, since taking $x^{\ast }\in core(Y)$ a.s.
satisfying $|x^{\ast }(N)-E(\tilde{c}_{R}^{T}(N))|\leq \phi _{T}$ for all $%
T\geq T^{\prime },$ we have%
\begin{equation*}
P[x^{\ast }(S)>Y_{S},\text{ for some }S\subset N]=0.
\end{equation*}

\medskip

Hence, we can conclude that it must exist $\widehat{x}\in \mathbb{R}^{n}$
such that $P[\widehat{x}\in core(\tilde{c}_{R}^{T},\varepsilon _{T})]=1$ for
all $T\geq T^{\prime }.$

\medskip

Second, by Theorem \ref{t:core}, any accumulation point $x^{T}\ $ of
Lehrer's $R_{2}$ repeated allocation process, at stage $T$, satisfies $%
P[x^{T}\in core(\tilde{c}_{R}^{T},\varepsilon _{T})]=1$.

\medskip

Then, we prove that for any accumulation point $\mathrm{x}$ of a sequence $%
\{x^{T}\}_{T\geq 1}$ such that $x^{T}\in core(\tilde{c}_{R}^{T},\varepsilon
_{T})$ a.s. for all $T>T^{\prime },$ one has that $\mathrm{x}\in core(Y)$
almost surely.

\medskip

Since any $\tilde{c}_{R}^{T}(S)\overset{a.s.}{\underset{T\rightarrow \infty }%
{\longrightarrow }}Y_{S}$ for all $S$. Thus, for any $\delta >0$, small
enough, there exists $T(\delta )$ such that for all $T>T(\delta )$: $\tilde{c%
}_{R}^{T}(S)<Y_{S}+\delta $ almost surely. Now, because $x^{T}\in core(%
\tilde{c}_{R}^{T},\varepsilon _{T})$ almost surely, it follows that:

\begin{equation*}
\begin{array}{l}
x^{T}(S)\leq \tilde{c}_{R}^{T}(S)+\varepsilon _{T}<Y_{S}+\delta +\varepsilon
_{T}, \; a.s., \\
\phi _{T}\geq |x^{T}(N)-E(\tilde{c}_{R}^{T}(N))|,%
\end{array}%
\end{equation*}
for all $T>T(\delta )$ and $S\subset N.$

\medskip

Hence, as $\delta \rightarrow 0$ we have that $T(\delta )\rightarrow \infty $
and the accumulation point $\mathrm{x}$ satisfies:
\begin{equation}
\mathrm{x}(S)\leq Y_{S}\mbox{ for all }S\subset N,\text{ and }|\mathrm{x}%
(N)-E(Y_{N})|\leq \phi _{Y},  \label{hello}
\end{equation}%
almost surely since $\varepsilon _{T}(S)\overset{a.s.}{\underset{%
T\rightarrow \infty }{\longrightarrow }}0$ for all $S\subseteq N$ and $\phi
_{T}\underset{T\rightarrow \infty }{\rightarrow }\phi _{Y}.$

\medskip

The sequence $\{\bar{a}^{R_{2}}(\tilde{c}_{R}^{T})(T)\}_{T\geq 1}$ satisfies
$\bar{a}^{R_{2}}(\tilde{c}_{R}^{T})(T)\in core(\tilde{c}_{R}^{T},\varepsilon
_{T})$ almost surely for all $T>T^{\prime }$. Hence, applying (\ref{hello})
any accumulation point of $\{\bar{a}^{R_{2}}(\tilde{c}_{R}^{T})(T)\}$
belongs to $core(Y)$ almost surely.
\end{proof}



\bigskip

\begin{thebibliography}{99}
\bibitem{AB99} \textsc{Anupindi, R. and Y. Bassok}, (1999).
\textquotedblleft Centralization of stocks: Retailers vs. Manufacturer",
\textit{Management Science} 45, 178-191

\bibitem{B56} \textsc{Blackwell, D.} (1956). \textquotedblleft An analog of
the MinMax Theorem for vector payoffs", \textit{Pacific J. of Math., 6, 1-8}.

\bibitem{B06} \textsc{Burer, S. and M. Dror} (2006). \textquotedblleft
Convex optimization of centralized inventory operation", \textit{submitted
for publication}.

\bibitem{CL89} \textsc{Chen, M.-S. and C.-T. Lin,} (1989). \textquotedblleft
Effects of Centralization on Expected Costs in a Multi-location Newsboy
Problem", \textit{Journal of the Operational Research Society} 40, 597-602.

\bibitem{E79} \textsc{Eppen, G. D.}, (1979). \textquotedblleft Effects of
centralization on expected costs in a multi-location newsboy problem",
\textit{Management Science} 25, 498-501.

\bibitem{F66} \textsc{Feller, W.}, (1966). An Introduction to Probability
Theory and its Applications, Vol. II, Wiley.

\bibitem{FPZ02} \textsc{Fernandez F. R., J. Puerto L. and M. J. Zafra},
(2002). \textquotedblleft Cores of Stochastic cooperative games", \textit{%
International Game Theory Review} 4, 3, 265-280.

\bibitem{G77} \textsc{Granot, D. }(1977). \textquotedblleft Cooperative
games in stochastic function form", Management Science 23, 621-630.


\bibitem{HD96} \textsc{Hartman, B. C. and M. Dror,} (1996). ``Cost
allocation in continuous review inventory models" \textit{Naval Research
Logistics Journal} 43, 549-561.

\bibitem{HD03} \textsc{Hartman, B. C. and M. Dror,} (2003). ``Optimizing
centralized inventory operations in a cooperative game theory setting",
\textit{IIE Transactions on Operations Engineering} 35, 243-257.

\bibitem{HD05} \textsc{Hartman, B. C. and M. Dror,} (2005). ``Allocation of
gains from inventory centralization in newsvendor environments", \textit{IIE
Transactions on Scheduling and Logistics} 37, 93-107.

\bibitem{HDS00} \textsc{Hartman, B. C., M. Dror, and M. Shaked,} (2000).
``Cores of inventory centralization games", \textit{Games and Economic
Behavior} 31, 26-49.


\bibitem{L02a} \textsc{Lehrer, E.,} (2002a), ``Allocation process in
cooperative games", \textit{International Journal of Game Theory} 31,
341-351.

\bibitem{L02b} \textsc{Lehrer, E.,} (2002b), ``Approachability in infinite
dimensional spaces", \textit{International Journal of Game Theory} 31,
253-268.


\bibitem{MSSS02} \textsc{Muller, A., M. Scarsini, and M. Shaked,} (2002).
``The newsvendor game has a nonempty core", \textit{Games and Economic
Behavior} 38, 118-126.

\bibitem{NA96} \textsc{Naurus J.A. and J.C. Anderson,} (1996). ``Rethinking
distribution", \textit{Harvard Business Review}, July-Aug 1996, 113-120.

\bibitem{P88} \textsc{Parlar, M.}, (1988). ``Game theoretic analysis of the
substitutable product inventory problem with random demands", \textit{Naval
Research Logistics} 35, 397-409.


\bibitem{RVZ95} \textsc{Ruiz, L.M., F. Valenciano, and J.M. Zarzuelo,}
(1998). \textquotedblleft The family of least square values for transferable
utility games", \textit{Games and Economic Behavior 24}, 109-130.




\bibitem{SFW05} \textsc{Slikker, M., J. Fransoo, and M. Wouters,} (2005).
\textquotedblleft Cooperation between multiple news-vendors with
transshipment", \textit{European J. of Operational Research} 167, 370-380.

\bibitem{Sui00} \textsc{Suijs J.,} (2000) ``Cooperative Decision-Making
under Risk''. Kluwer Academic Publishers, Boston.



\bibitem{T06} \textsc{Timmer, J.,} (2006). \textquotedblleft The compromise
value for cooperative games with random payoffs", \textit{Mathematical
Methods of Operations Research} 64, 95-106.

\bibitem{TBT03} \textsc{Timmer, J., P. Borm, and S. Tijs,} (2003).
\textquotedblleft On three Shapley-like solutions for cooperative games with
random payoffs", \textit{International Journal of Game Theory} 32, 595-613.

\bibitem{TBT05} \textsc{Timmer, J., P. Borm, and S. Tijs,} (2005).
\textquotedblleft Convexity in stochastic cooperative situations", \textit{%
International Game Theory Review 7}, 25-42.
\end{thebibliography}
\end{document}